\documentclass[12pt,preprint]{aastex}

\begin{document}
    
\title{Multiple inverse Compton scatterings and the blazar sequence}

\author{C.-I. Bj\"ornsson\altaffilmark{1}}
\altaffiltext{1}{Department of Astronomy, AlbaNova University Center, Stockholm University, SE--106~91 Stockholm, Sweden.}
\email{bjornsson@astro.su.se}

\begin{abstract}
The high frequency component in blazars is thought to be due to inverse Compton scattered radiation. Recent observations by {\it Fermi}-LAT are used to evaluate the details of the scattering process. A comparison is made between the usually assumed single scattering scenario and one in which multiple scatterings are energetically important. In the latter case, most of the radiation is emitted in the Klein-Nishina limit. It is argued that several of the observed correlations defining the blazar sequence are most easily understood in a multiple scattering scenario. Observations indicate also that, in such a scenario, the blazar sequence is primarily governed by the energy density of relativistic electrons rather than that of the seed photons. The pronounced X-ray minimum in the spectral energy distribution often observed in the most luminous blazars is discussed. It is shown how this feature can be accounted for in a multiple scattering scenario by an extension of standard one-zone models.
\end{abstract}


\section{Introduction}

The spectral energy distribution in blazars exhibits two distinct components. It was recognized early on that the low frequency component is almost certainly due to synchrotron radiation. The high frequency component is likely to be inverse Compton (IC) scattered radiation, although various hadronic origins have also been considered \citep[e.g.,][]{man93,m/p01}. \citet{p/g95} suggested a classification of blazars based on the frequency where the synchrotron component peaks. The distribution of this peak frequency showed a rough bimodal distribution, which gave rise to three classes; two classes corresponding to the two largest groups of blazars with peak frequencies in the infrared and in the UV/X-ray, respectively, and a third, smaller class encompassing those with intermediate peak frequencies. This rough classification scheme still remains useful although the detailed criteria for each class, as well as the nomenclature, have changed. In this paper the nomenclature introduced by \citet{abd09b} is followed. The Low Synchrotron Peaked blazar (LSP) class includes both flat spectrum radio quasars (FSRQ) and low frequency peaked BL Lac objects, while the other two classes, Intermediate Synchrotron Peaked blazars (ISP) and High Synchrotron Peaked blazars (HSP), correspond roughly to intermediate and high frequency peaked BL Lac objects.

With the advent of more observational information about the high frequency component, this classification scheme was developed into a blazar sequence by \citet{fos98}. It was found that the properties of the IC-component followed a trend reflecting those of the synchrotron component; a lower peak frequency of the synchrotron component corresponds to a higher luminosity and a lower peak frequency of the IC-component. This could be understood as a result of increased cooling in the more luminous blazars. The more extensive and detailed observations by {\it Fermi}-LAT have shown that these trends not only connect the three classes of blazars but actually constitute smooth and continuous relations between the properties of the IC-component and the peak frequency of the synchrotron component \citep{abd10b}. Furthermore, {\it Fermi}-LAT has revealed an additional correlation, which shows that more luminous blazars tend to have steeper spectra in the LAT frequency range \citep{abd09a,abd10a}.
 
It is usually agreed that the electrons producing the IC-component are responsible also for at least the high frequency part of the synchrotron component \citep[e.g.,][]{mar92,d/s93,sik94,b/m96}. The origin of the seed photons for the IC-component is somewhat more uncertain; they could either be external photons or the intrinsically produced synchrotron photons (SSC-models). Since there is a rough correlation between emission line strength and luminosity of the IC-component, it has been suggested \citep{ghi98} that external photons dominate the cooling in the most luminous blazars, while the least luminous blazars correspond to the SSC-case. It is normally assumed that the dominant contribution comes from first order inverse Compton scattering \citep[e.g.,][]{g/t09, sik09}. The spectral properties are then determined mainly by the energy distribution of the injected relativistic electrons. Another possibility is that the source properties are such that the IC-component is due to multiple inverse Compton scatterings (MIC). In this case, its spectral properties depend to a large extent also on the scattering process itself. In addition to different spectral characteristics, time variations are expected to have properties which differ between MIC-models and first order scattering models; this is particular true for correlations between flux variations in the synchrotron and IC components.

In \citet[][henceforth BA00]{b/a00} MIC-models were discussed in connection with the observed properties of the radio/mm outbursts in blazars. Motivated by the {\it Fermi}-LAT observations of blazars, the present paper extends this discussion to the high frequency radiation. Guided by the numerical results in BA00, an analytical description of MIC-models is developed in \S\,2. It is emphasized that the origin of the seed photons only marginally affects the results. The consequences for the synchrotron component is briefly discussed in \S\,3, in particular the effects of cooling on the synchrotron self-absorption frequency. The most luminous blazars often exhibit a distinct X-ray minimum in their spectral energy distribution, which is hard to account for in one-zone MIC-models  \citep[e.g.,][]{b/m96}. The extensions of standard one-zone models needed to accomodate such minima in MIC-models are discussed in \S\,4.  In MIC-models, the relativistic electrons radiate most of their energy in the Klein-Nishina limit. Hence, with a typical bulk Lorentz factor $\Gamma \sim 10$ for the emission region in blazars, the photons detected by {\it Fermi}-LAT should be up-scattered from a frequency range overlapping that of {\it Swift}-BAT. \citet[][henceforth S09, see also \citealt{abd09b}]{sam09} have discussed the properties of a sample of blazars detected both by {\it Fermi}-LAT and {\it Swift}-BAT. This is taken as the starting point in \S\,5 for an evaluation of how MIC-models and first order scattering models fare in a comparison with observations. It is argued that several observed properties are more straightforwardly accounted for by a multiple scattering scenario than by one where only a single scattering occurs. This has implications for the origin of the blazar sequence, since in a MIC-scenario observations indicate that this sequence is not primarily due to variations in the energy density of seed photons but rather to the energy density of relativistic electrons. The main conclusions are collected in \S\,6.

\section{The inverse Compton spectrum} \label{sect2}
Zel'dovich \citep[see][]{r/l79,poz76} originally proposed that the spectrum resulting from multiple inverse Compton scatterings by thermal/mono-energetic electrons in the Thomson limit can be approximated globally by a power-law.
The usefulness of this approximation for describing the multiply inverse Compton scattered radiation depends mainly on the number of orders that can be accomodated before the Klein-Nishina limit sets in. This in turn is set by a combination of the frequency of the seed photons and the energy of the electrons, which give the dominant contribution to the inverse Compton flux. For a power-law distribution of relativistic electrons the latter energy needs to be determined in a self-consistent way.

\subsection{Contributions to the inverse Compton flux} \label{sect2a}

Consider a power-law distribution of flux ($f(\nu)$) with spectral index $\alpha$ ($f(\rm \nu)\propto \nu^{-\alpha}$) due to multiple inverse Compton scatterings by an electron distribution $N(\gamma)=K\gamma^{-p}$ with $p>2$ and $\gamma\ge\gamma_{\rm min}$, where $\gamma$ is the Lorentz factor of the relativistic electrons. Assume further that $\alpha<1$ so that cooling occurs in the Klein-Nishina limit, i.e., electrons cool on photons with frequency $\nu\propto\gamma^{-1}$. Let the electrons be uniformly injected into a region of size $r_{\rm o}$ from which they escape with speed $v_{\rm esc}$. Define $\gamma_{\rm c}$ such that its cooling time $t_{\rm c}(\gamma=\gamma_{\rm c})\equiv r_{\rm o}/v_{\rm esc}$. The cooling time for electrons with $\gamma>\gamma_{\rm c}$ is then given by
\begin{equation}
	\frac{t_{\rm c}(\gamma>\gamma_{\rm c})}{t_{\rm c}(\gamma_{\rm c})}=\left(\frac{\gamma_{\rm c}}		{\gamma}\right)^{\alpha}.
	\label{eq:1.1}
\end{equation}

In a steady state situation, the main contribution of photons to the energy density $U_{\rm ph}(\nu)$($\propto \nu f(\nu)$) at frequency $\nu$ comes from electrons with $\gamma$ determined by (cf. BA00)
\begin{eqnarray}
	\frac{U_{\rm ph}(\nu,\gamma)}{U_{\rm ph}(\nu,\gamma_{\rm c})}&=&\left(\frac{\gamma} 
	{\gamma_{\rm c}} \right)^{3-p}\frac{r(\gamma)}{r_{\rm o}}\left(\frac{\gamma_{\rm c}}{\gamma}\right)^{2(1-
	\alpha)} \nonumber\\
	&=&\frac{r(\gamma)}{r_{\rm o}}\left(\frac{\gamma_{\rm c}}{\gamma}\right)^{p-1-2\alpha}.
	\label{eq:1.2}
\end{eqnarray}
The factor $\propto\gamma^{3-p}$ corresponds to the energy gain ($\gamma^2$) per scattering times the density of electrons with Lorentz factor $\gamma$ ($\propto\gamma^{1-p}$), while the factor $\propto\gamma^{-2(1-\alpha)}$ takes into account the fact the electrons with different Lorentz factors scatter on different photons ($\nu\propto\gamma^{-2}$) and, hence, different energy densities. The cooling distance $r(\gamma)<r_{\rm o}$ for $\gamma>\gamma_{\rm c}$. For $\gamma_{\rm c}>\gamma_{\min}$, there are two cases.

$\alpha>(p-1)/2$:  If $\gamma>\gamma_{\rm c}$ then $r(\gamma)/r_{\rm o}=(\gamma_{\rm c}/\gamma)^{\alpha}$ and the ratio in equation (\ref{eq:1.2}) is $(\gamma_{\rm c}/\gamma)^{p-1-\alpha}$ which is smaller than unity, since $p>2$. Hence, electrons with $\gamma>\gamma_{\rm c}$ cannot be the main contributors to the flux in the inverse Compton component. For $\gamma<\gamma_{\rm c}$, $r(\gamma)=r_{\rm o}$ and the ratio in equation (\ref{eq:1.2}) is given by $(\gamma_{\rm c}/\gamma)^{p-1-2\alpha}$, which is also smaller than unity, since $\alpha>(p-1)/2$. Therefore, the flux in the inverse Compton component is dominated by electrons with $\gamma=\gamma_{\rm c}$.

$\alpha<(p-1)/2$: For $\gamma>\gamma_{\rm c}$ the result is the same as above. However, when $\alpha<(p-1)/2$ the ratio in equation (\ref{eq:1.2}) is larger than unity for $\gamma<\gamma_{\rm c}$. Therefore, in this case, the flux in the inverse Compton component is dominated by the electrons with lowest energies, i.e., $\gamma=\gamma_{\rm min}$. This case also includes $p>3$, since $\alpha<1$ is a requirement for the present discussion.

When $\gamma_{\rm c}<\gamma_{\rm min}$, cooling is important for all the electrons. It is straightforward to show with arguments similar to those above that in this case the inverse Compton flux is dominated by electrons with $\gamma=\gamma_{\rm min}$ independent of the value of $\alpha$.

\subsection{Cooling characteristics} \label{sect2b}
The results in \S~\ref{sect2a} indicate that the spectral shape is intimately connected to the cooling characteristics of the source, i.e., the value of $\gamma_{\rm c}$. The cooling time for an electron with Lorentz factor $\gamma$ is
\begin{equation}
	t_{\rm c}(\gamma)=\frac{3mc}{4\sigma_{\rm T}U_{\rm ph}(\nu)\gamma},
	\label{eq:1.3}
\end{equation}
where $\sigma_{\rm T}$ is the Thomson cross-section. Since cooling occurs in the Klein-Nishina limit, the frequency to use in equation (\ref{eq:1.3}) is $\nu\propto\gamma^{-1}$. The cooling characteristics differ somewhat in the three cases discussed above.

$\alpha>(p-1)/2$: $U_{\rm ph}(\nu)$ is determined by electrons with $\gamma=\gamma_{\rm c}$. It can be expressed as 
\begin{equation}
	U_{\rm ph}(\nu)=\left(\frac{\nu}{\nu_{\rm max}}\right)^{1-\alpha}U_{\rm ph,o},
	\label{eq:1.4}
\end{equation}
where $\nu_{\rm max}=\gamma_{\rm c}mc^2/h$ and $U_{\rm ph,o}$ is the flux resulting from the complete cooling of electrons with Lorentz factor $\gamma_{\rm c}$,
\begin{equation}
	\frac{cU_{\rm ph,o}}{4}=v_{\rm esc}\epsilon_{\rm e}(\gamma_{\rm c})U_{\rm e}.
	\label{eq:1.5}
\end{equation}
Here, $U_{\rm e}$ is the total energy density of electrons injected into the source and $\epsilon_{\rm e}(\gamma_{\rm c})=(\gamma_{\rm c}/\gamma_{\rm min})^{2-p}$ is the fraction thereof that has energies around $\gamma=\gamma_{\rm c}$. 

Since $t_{\rm c}(\gamma_{\rm c})\equiv r_{\rm o}/v_{\rm esc}$, equation (\ref{eq:1.3}) implies
\begin{equation}
	\tau_{\rm o}=\gamma_{\rm c}^{p-1-2\alpha},
	\label{eq:1.8}
\end{equation}
which relates the values of $\alpha$ and $\gamma_{\rm c}$. Here,
\begin{equation}
	\tau_{\rm o}=\frac{8(p-1)}{3(p-2)}\gamma_{\rm min}^{p-1}n_{\rm e}\sigma_{\rm T} r_{\rm o},
	\label{eq:1.7}
\end{equation}
where $n_{\rm e}$ is the total number density of electrons. Hence, $\tau_{\rm o}$ is roughly the Thomson scattering depth ($\tau_{\rm T}$) for an electron distribution with $\gamma_{\rm min}=1$ ( i.e., $\tau_{\rm o} \approx \tau_{\rm T} \gamma_{\rm min}^{p-1}$) . For consistency, this implies $\tau_{\rm o}<1$, since $\alpha>(p-1)/2$.
 
$\alpha<(p-1)/2$: $U_{\rm ph}(\nu)$ is determined by electrons with $\gamma=\gamma_{\rm min}$. The difference is now that $\nu_{\rm max}=\gamma_{\rm min}mc^2/h$ and that $\epsilon_{\rm e}(\gamma_{\rm c})$ is substituted by $\left(\gamma_{\rm min}/\gamma_{\rm c}\right)^{\alpha}$, which is the fraction of the energy that electrons with $\gamma=\gamma_{\rm min}$ lose before escaping. In analogy with the analysis above, this leads to
\begin{equation}
	\tau_{\rm o}=\gamma_{\rm min}^{p-1-2\alpha}.
	\label{eq:1.9}
\end{equation}
Again, consistency implies $\tau_{\rm o}>1$, since $\alpha<(p-1)/2$. Hence, if $\tau_{\rm o}=1$ occurs for $\gamma_{\rm c}>\gamma_{\rm min}$, $\alpha=(p-1)/2$ at this instance. Furthermore, as can be seen from the definition of $\tau_{\rm o}$ (cf. eq. [\ref{eq:1.7}]), equation (\ref{eq:1.9}) is equivalent to $\tau_{\rm T} \approx \gamma_{\rm min}^{-2\alpha}$, which is the result originally derived by Zel'dovich. The reason is that in this limit, electrons with $\gamma > \gamma_{\rm min}$ contribute only marginally to the inverse Compton spectrum and, hence, the electron distribution corresponds effectively to a monoenergetic distribution. This also accounts for the fact that $\alpha $ is independent of $\gamma_{\rm c}$.  

When $\gamma_{\rm c}<\gamma_{\rm min}$, $U_{\rm ph}(\nu)$ is determined by electrons with $\gamma=\gamma_{\rm min}$. However, in contrast with the previous case, electrons with $\gamma=\gamma_{\rm min}$ lose all the energy and, hence, the factor $\left(\gamma_{\rm min}/\gamma_{\rm c}\right)^{\alpha}$ is substituted by unity, which leads to
\begin{equation}
	\tau_{\rm o}=\gamma_{\rm min}^{p-1-2\alpha}\left(\frac{\gamma_{\rm min}}{\gamma_{\rm c}}			\right)^{\alpha}.
	\label{eq:1.10}
\end{equation}
Physically, the factor $\left(\gamma_{\rm min}/\gamma_{\rm c}\right)^{\alpha}$ accounts for the decrease in the column density of electrons with $\gamma = \gamma_{\rm min}$ caused by cooling.

The value of $\alpha$ describes the upper envelope connecting the different Compton orders and is determined by electrons of a given energy (i.e., $\gamma_{\rm min}$ or $\gamma_{\rm c}$) . Hence, the local spectral index can deviate substantially from $\alpha$, in particular, for large values of $\gamma_{\rm min}$ or $\gamma_{\rm c}$. However, as discussed in BA00, a power-law distribution of electron energies considerably smoothens the spectrum as compared to a thermal/mono-energetic distribution. A distinct feature of a power-law distribution is that the energy of the electrons determining the value of $\alpha$ changes rapidly from $\gamma_{\rm c}$ to $\gamma_{\rm min}$ (or vice versa) as the value of $\tau_{\rm o}$ goes through unity. Although the value of $\alpha$ changes smoothly in this transition, the local spectral index can change rapidly, in particular for cases with $\gamma_{\rm c} \gg\gamma_{\rm min}$ at this instance. This change in spectral character from a "bumpy" to a smooth spectrum as $\gamma_{\rm c}$ is replaced by $\gamma_{\rm min}$ as the main contributor to the inverse Compton emission is evident in the spectra presented in BA00.

These expressions for $\alpha$ and $\gamma_{\rm c}$ for a given value of $\tau_{\rm o}$ depend only on the assumption of cooling in the Klein-Nishina limit. In order to relate the values for $\alpha$ and $\gamma_{\rm c}$ to the source properties, the origin of the seed photons needs to be considered. Two different scenarios are usually envisaged; external photons, whose properties are unrelated to those of the relativistic electrons, and synchrotron photons produced by the relativistic electrons themselves (SSC-models).  

\subsection{The seed photons} \label{sect2c}
When the seed photons are due to synchrotron radiation from the same electrons producing the inverse Compton flux, a particular simple description is possible. This is due to the fact that synchrotron radiation can be regarded as inverse Compton scattering of the virtual photons associated with the magnetic field. Hence, the second equation needed to obtain separate values for $\alpha$ and $\gamma_{\rm c}$ is straightforward to derive.

$\alpha>(p-1)/2$: The expression relating the synchrotron and inverse Compton components is
\begin{equation}
	U_{\rm B}\left(\frac{\gamma_{\rm c}mc^2}{h\nu_{\rm B}}\right)^{1-\alpha}=U_{\rm ph,o},
	\label{eq:1.11}
\end{equation}
where $U_{\rm ph,o}$ is given by equation (\ref{eq:1.5}). The energy density of the magnetic field is denoted by $U_{\rm B}$ and the cyclotron frequency by $\nu_{\rm B}$. This yields  
\begin{equation}
	\left(\frac{\gamma_{\rm c}mc^2}{h\nu_{\rm B}}\right)^{1-\alpha}=\left(\frac{\gamma_{\rm c}}
	{\gamma_{\rm min}}\right)^{2-p}\frac{\epsilon_{\rm e}}{\epsilon_{\rm B}},
	\label{eq:1.12}
\end{equation}
where $\epsilon_{\rm e}/\epsilon_{\rm B}\equiv 4v_{\rm esc}U_{\rm e}/cU_{\rm B}$.

In order to single out the density dependence, it is useful to write
\begin{equation}
	\frac{\epsilon_{\rm e}}{\epsilon_{\rm B}}\equiv \epsilon_{\rm eBo}\tau_{\rm o},
	\label{eq:1.13}
\end{equation}
so that $\epsilon_{\rm eBo}$ is the value of $\epsilon_{\rm e}/\epsilon_{\rm B}$ for $\tau_{\rm o}=1$. Convenient expressions for $\alpha$ and $\gamma_{\rm c}$ are obtained by evaluating equation (\ref{eq:1.12}) for $\tau_{\rm o}=1$ (i.e., $\alpha=(p-1)/2$), which yields
\begin{equation}
	\left(\frac{\gamma_{\rm co}}{\gamma_{\rm min}}\right)^{(p-1)/2} = \left(
	\frac{\gamma_{\rm min}mc^2}{h\nu_{\rm B}}\right)^{(p-3)/2}
	\epsilon_{\rm eBo},
	\label{eq:1.14}
\end{equation}
This alternative parameter ($\gamma_{\rm co}$) is the value of $\gamma_{\rm c}$ corresponding to $\tau_{\rm o}=1$. Hence, the synchrotron-inverse Compton spectrum is basically described by two parameters: The value of $\tau_{\rm o}$, which depends on the column density of electrons but not the magnetic field and $\gamma_{\rm co}$, whose value includes the magnetic field but not the column density of electrons. In the external photon scenario, the same discussion applies, except that the frequency and energy density of the seed photons are not {\it a priori} related and, hence, the value of $\gamma_{\rm co}$ depends on two parameters instead of just one as in the SSC-case. 

With the use of equations (\ref{eq:1.8}) and (\ref{eq:1.11}), separate expressions for $\alpha$ and $\gamma_{\rm c}$ can be derived in the form of two second order equations (cf. BA00). However, in order to highlight a few characteristics of $\gamma_{\rm c}$, it is more convenient to express them as
\begin{equation}
	y^{\frac{p-1-2\alpha}{2}} = \tau_{\rm o}^{\frac{-\alpha}{p-1-2\alpha}}\gamma_{\rm co}^{\frac{p-1}{2}}
	\label{eq:1.15}
\end{equation}
and
\begin{equation}
	\frac{\gamma_{\rm c}}{\gamma_{\rm co}} = \tau_{\rm o}^{\frac{1}{p-1}\left(1-\frac{\ln y}{\ln 
	\gamma_{\rm c}}\right)},
	\label{eq:1.16}
\end{equation}
where $y \equiv mc^2/h \nu_{\rm B}$ has been introduced. Since for blazars the value of $y$ is likely to be very large (e.g., $y \sim 10^{14}$ for $B \sim 1$), it is seen from equation (\ref{eq:1.16}) that the value for $\gamma_{\rm c}$ is expected to be quite sensitive to that of $\tau_{\rm o}$. Furthermore, as the value of  $\tau_{\rm o}$ increases, the decrease of $\gamma_{\rm c}$ accelerates, due to the $1/\ln  \gamma_{\rm c}$ factor in the exponent. Both of these properties derive from the fact that the effective number of inverse Compton orders is proportional to $\ln y/ \ln \gamma_{\rm c}$.

In the external photon scenario, equations (\ref{eq:1.15}) and (\ref{eq:1.16}) completely describe the relation between the source properties and the values of $\alpha$ and $\gamma_{\rm c}$. This is not quite the case for the SSC-scenario, since it was tacitly assumed in equation (\ref{eq:1.11}) that $\gamma_{\rm c} > \gamma_{\rm abs}$, where $\gamma_{\rm abs}$ is the Lorentz factor corresponding to the synchrotron self-absorption frequency ($\nu_{\rm abs}$). When $\gamma_{\rm c} < \gamma_{\rm abs}$, the energy density of synchrotron photons peaks at $\nu_{\rm abs}$ rather than the cooling frequency ($\nu_{\rm c} = \gamma_{\rm c}^2 \nu_{\rm B}$). Hence, the same analysis as above can be made except that two substitutions need to be made: $U_{\rm B} \rightarrow U_{\rm ph}(\nu_{\rm abs})$ and $\nu_{\rm B} \rightarrow \nu_{\rm abs} = \gamma_{\rm abs}^2 \nu_{\rm B}$. Since
\begin{equation}
	U_{\rm ph}(\nu_{\rm abs}) = U_{\rm B} \tau_{\rm o} \gamma_{\rm abs}^{3-p} \left(
	\frac{\gamma_{\rm c}}{ \gamma_{\rm abs}}\right)^{\alpha},
	\label{eq:1.17}
\end{equation}
where the last factor on the RHS accounts for the decreasing column density of electrons with $\gamma = \gamma_{\rm abs}$ due to cooling, the expression corresponding to equation (\ref{eq:1.16}) is 
\begin{equation}
	\frac{\gamma_{\rm c}}{\gamma_{\rm co}}\left(\frac{\gamma_{\rm abso}}{\gamma_{\rm abs}}
	\right)^{\frac{1}{2}}=
	\tau_{\rm o}^{\frac{1}{p-1}\left(\frac{\ln \gamma_{\rm abs} -\ln y}{2\ln \gamma_{\rm c}}\right)}.
	\label{eq:1.18}
\end{equation}
When $\tau_{\rm o} = 1$ occurs for $\gamma_{\rm c} < \gamma_{\rm abs}$, $\gamma_{\rm abso}$ is the value of $\gamma_{\rm abs}$ at $\tau_{\rm o} = 1$. In this case, separate values for $\gamma_{\rm co}$ and $\gamma_{\rm abso}$ cannot be obtained; however, their combination, which occurs in equation (\ref{eq:1.18}), is given by an expression similar to that in equation (\ref{eq:1.14})
\begin{equation}
	\left(\frac{\gamma_{\rm co}}{\gamma_{\rm abso}^{1/2}}\right)^{(p-1)} = \gamma_{\rm 
	min}^{p-2}\left(\frac{mc^2}{h\nu_{\rm B}}\right)^{(p-3)/2}
	\epsilon_{\rm eBo},
	\label{eq:1.19}
\end{equation}
The main difference between equation (\ref{eq:1.16}) and  equation (\ref{eq:1.18}) is that the exponent of the RHS-side is roughly a factor of two smaller for the latter as compared to the former. Formally the reason for this is that when $\tau_{\rm o} = 1$ occurs for $\gamma_{\rm c} > \gamma_{\rm abs}$, $\gamma_{\rm abso} \rightarrow \gamma_{\rm co}$ and equation (\ref{eq:1.19}) becomes identical to equation (\ref{eq:1.14}). Hence, the variation of $\gamma_{\rm c}$ with $\tau_{\rm o}$ slows down considerably for  $\gamma_{\rm c} < \gamma_{\rm abs}$. The physical reason for this is that the energy density of synchrotron seed photons varies much slower in this regime. This conclusion hinges on the assumption that the value of $\gamma_{\rm abs}$ varies more slowly than $\gamma_{\rm c}$. It is shown in the next section that this is indeed the case. 
   
$\alpha<(p-1)/2$: In this regime, $\alpha$ is obtained directly from equation (\ref{eq:1.9}). The variation of $\gamma_{\rm c}$  is quite similar to that for $\alpha>(p-1)/2$ and expressions corresponding to equations (\ref{eq:1.16}) and (\ref{eq:1.18}) can be derived. The main difference is that $\gamma_{\rm c}$ is replaced by $\gamma_{\rm min}$ in the exponent of $\tau_{\rm o}$, which is due to the fact that the main contribution to the inverse Compton flux now comes from electrons with $\gamma =\gamma_{\rm min}$. These results also remain valid for the regime $\gamma_{\rm c} <\gamma_{\rm min}$.

\section{The synchrotron spectrum}\label{sect3}
In order to relate the properties of the synchrotron component to those of the inverse Compton component, the value of $\gamma_{\rm abs}$ needs to be determined self-consistently. Since the main parameter in the multiple inverse Compton scenario is $\tau_{\rm o}$, in this section attention is restricted to the case when only the column density of electrons varies.

When $\gamma_{\rm c} > \gamma_{\rm abs}$, standard synchrotron theory gives \citep[e.g.,][]{pac70} 
\begin{equation}
	\gamma_{\rm abs}\propto \tau_{\rm o}^{\frac{1}{(4+p)}}. 	
	\label{eq:2.1}
\end{equation}
Cooling affects the absorption properties for $\gamma_{\rm c} < \gamma_{\rm abs}$, since the column density of electrons with Lorentz factor $ \gamma_{\rm abs}$ varies as $\left(\gamma_{\rm c} / \gamma_{\rm abs}\right)^{\alpha}$. Hence, the expression for $\gamma_{\rm abs}$ in this case is obtained by letting $\tau_{\rm o} \rightarrow \tau_{\rm o} \left(\gamma_{\rm c} / \gamma_{\rm abs}\right)^{\alpha}$ in equation (\ref{eq:2.1}), which leads to
\begin{equation}
	\gamma_{\rm abs}\propto \left(\tau_{\rm o}\gamma_{\rm c}^{\alpha}\right)^{\frac{1}{4+\alpha +p}}. 	
	\label{eq:2.2}
\end{equation}
The expression for $\gamma_{\rm c}$ to use in equation (\ref{eq:2.2}) is that given either in equation (\ref{eq:1.16}) for the external photon case or equation  (\ref{eq:1.18}) for the SSC-case. For external seed photons, this yields
\begin{equation}
	\gamma_{\rm abs} \propto \tau_{\rm o}^{\frac{(1-\ln y / 2 ln \gamma_{\rm c})}{4+\alpha+p}},
	\label{eq:2.3}
\end{equation}
where equation (\ref{eq:1.8}) has been used. As mentioned above, for $\alpha < (p-1)/2$ the expression for $\gamma_{\rm abs}$ is quite similar to the one for $\alpha > (p-1)/2$, except that $\gamma_{\rm c}$ is replaced by $\gamma_{\rm min}$. In the SSC-case, the variation of $\gamma_{\rm c}$ with $\tau_{\rm o}$ is considerably slower than for external photons (cf. eqns. [\ref{eq:1.16}] and [\ref{eq:1.18}]) and, hence, the exponent in equation (\ref{eq:2.3}) is correspondingly smaller. However, the actual value of the exponent is likely to be rather similar in the two cases, since the value of $y$ is expected to be much larger in the SSC-case as compared to the external photon case (e.g., $y\sim 10^{14}$ for $B \sim 1$ as compared to $y\sim 10^{5}$ for optical seed photons). 

With $\gamma_{\rm c} \lesssim \gamma_{\rm abs} \sim 10^2$, it is likely that $\ln y > 2\ln \gamma_{\rm c}$ independent of the origin of the seed photons. The dependence of  $\gamma_{\rm abs}$ on $\tau_{\rm o}$ would then reverse as compared to the case $\gamma_{\rm c} > \gamma_{\rm abs}$; instead of having a value of $\gamma_{\rm abs}$ increasing with $\tau_{\rm o}$, it now decreases.

\section{Connection to blazars}\label{sect4}

In a MIC-scenario, most of the main characteristics of the inverse Compton component are quite insensitive to the origin of the seed photons and are instead the result of the scattering process itself. An example is the expected spectral correlations between the part determined by $\alpha$ (i.e., mainly X-ray) and the spectral range where the emission occurs in the Klein-Nishina limit (i.e., mainly gamma-ray).

These spectral correlations can be expressed in terms of the spectral breaks that occur at frequencies above the X-ray regime (see BA00). There will be one or two breaks depending on the value of $\alpha$. For $\alpha > (p-1)/2$, the X-ray spectrum is determined by $\gamma_{\rm c}$ and extends to $h\nu \approx \gamma_{\rm c} mc^2$. Above this frequency, cooling is important and the spectral index is given by $p-1$ \citep{b/g70}; hence, the magnitude of this spectral break is $\Delta \alpha = p-1-\alpha$. This is also the frequency where the value of $\nu f(\nu)$ peaks. For $\alpha < (p-1)/2$, this single break splits into two. The X-ray spectrum is now determined by $\gamma_{\rm min}$ and extends to $h\nu \approx \gamma_{\rm min} mc^2$. Above this frequency, emission occurs in the Klein-Nishina limit implying that the emitted frequency is proportional to electron energy. Since the electrons lose only a fraction of their energy ($\propto \gamma^{\alpha}$), the spectral index is given by $p-1-\alpha$, which corresponds to a spectral break of magnitude $\Delta \alpha_{\rm 1} = p-1-2 \alpha$. A second, cooling break at $h\nu \approx \gamma_{\rm c} mc^2$ has magnitude $\Delta \alpha_{\rm 2} = \alpha$. The peak in $\nu f(\nu)$ occurs at the first break when $\alpha < p-2$ and at the second break when $\alpha > p-2$; the latter requires $p<3$. The two breaks merge again into one for $\gamma_{\rm c} < \gamma_{\rm min}$. These spectral properties are summarized in Figure\,\ref{fig1}, which shows schematically how the inverse Compton scattered part of the spectrum changes as the column density of electrons (i.e., $\tau_{\rm o}$) increases. It may be noticed that of the three possible frequency ranges with different spectral indices only the one at the highest frequencies is determined exclusively by the energy distribution of the injected electrons (i.e., $p$). The other two spectral indices can vary continuously without any change in $p$. Such variations are instead caused by changes in the electron density and are correlated through the value of $\alpha$.

The source properties of blazars deduced from observations in a MIC-scenario differ in some crucial aspects from those in a single scattering scenario; however, several remain the same. In a MIC-scenario, the value of $\gamma_{\rm min}$ should be small to assure a smooth inverse Compton spectrum; from BA00 this implies $\gamma_{\rm min} \lesssim 10$. Similar values are also suggested for single scattering of photons from the broad line region, since the X-ray spectra measured by {\it Swift}-BAT is on average somewhat too soft to be due to the cooling tail of higher energy electrons \citep[e.g.,][]{sam09,abd09b}. For the most likely site of the emission region in a single scattering scenario, cooling on external photons leads to $\gamma_{\rm c} \lesssim 10^2$ in the most luminous blazars (LSPs)  \citep[e.g.,][]{g/t09,sik09}. This is also similar to the MIC-case and implies that the peak in the spectral energy distribution of the synchrotron component occurs close to the self-absorption frequency. 

The main difference between the two scenarios concerns the value of $\epsilon_{\rm e}/\epsilon_{\rm B}$. Single scattering models are usually consistent with rough equipartition between the energy densities in electrons and magnetic field (i.e., $\epsilon_{\rm e} \sim \epsilon_{\rm B}$), while multiple scattering models require $\epsilon_{\rm e} \gg \epsilon_{\rm B}$ (cf. eq. [\ref{eq:1.14}]). Furthermore, the smooth power-law X-ray spectra in LSPs indicate $\tau_{\rm o} \gtrsim 1$ in MIC-models in order for this radiation to be produced by the lowest energy electrons (cf. eq. [\ref{eq:1.9}]).

The distinct minimum in the spectral energy distribution occurring at the transition between the synchrotron and inverse Compton components, which is particularly pronounced in LSPs, is not consistent with a strictly homogeneous one-zone MIC-model. The low value of $\gamma_{\rm min}$ needed to produce the observed smooth power-law in the X-ray regime would also fill in this minimum by inverse Compton scattering of the synchrotron component \citep[e.g.,][]{b/m96}. However, any one-zone synchrotron/Compton model have problems accounting for some of the main characteristics of LSP; for example, although correlated time variations are sometimes seen  \citep[e.g.,][]{mar08,abd10c} between the optical and the gamma frequency ranges, this is not generally the case.

The similar value of the magnetic field $B$ in single and multiple scattering scenarios have two implications: (i) The deduced location of the emission site should be roughly the same. (ii) In LSPs most of the radiation is emitted by cooling electrons. Hence, the much larger value of the energy density of electrons in a MIC-model as compared to a single scattering model necessitates a correspondingly smaller emission volume, i.e., $r_{\rm o} \ll R$, where $R$ is the size of the available emission region. This suggests a possible extension of the one-zone MIC model that can account for the minimum in the spectral energy distribution.

For simplicity, assume that the values of $B$ as well as $v_{\rm esc}$ are the same throughout the emission region, including the smaller injection region. The uniform injection of relativistic electrons inside $r_{\rm o}$ implies that the fraction of electrons with $\gamma > \gamma_{\rm c}$, which escapes and fills the whole emission region $R$, is proportional to their cooling time. The contributions to the optically thin synchrotron flux is determined by the average time spent by an electron in the two regions. With no additional cooling outside the injection region, the shape of the synchrotron spectrum emitted by the escaping electrons is, therefore, similar to that inside $r_{\rm o}$, while the flux is a factor $R/r_{\rm o}$ larger. Although cooling on the inverse Compton component is less efficient by a factor $R/r_{\rm o}$ outside as compared to inside the injection region, cooling may steepen the synchrotron spectrum from the escaping electrons at higher frequencies. A further contribution could come from synchrotron cooling.  Hence, the smaller injection region contributes most of the inverse Compton flux, while the synchrotron emission comes mainly from the emission region as a whole. 

Within the injection region, the amplitude between consecutive Compton orders is roughly $\gamma_{\rm min}^2 \tau_{\rm T} \sim \gamma_{\rm min}^{2(1-\alpha)}$ (cf. eq. [\ref{eq:1.9}]). Outside the injection region this amplitude is smaller by a factor $R/r_{\rm o}$ due to the smaller column density of electrons. The small value of $\gamma_{\rm c}$ expected in LSPs causes the inverse Compton scattering also outside the injection region to be dominated by the low energy end of the electron distribution. The minimum in the spectral energy distribution observed in LSPs is typically a factor $10$ below the peak of the synchrotron component. In order to make MIC-models compatible with such minima, $\gamma_{\rm min}^{2(1-\alpha)} r_{\rm o} / R \lesssim 10^{-1/n}$ is required, where $n$ is the Compton order corresponding to the frequency of the minimum. Since $\gamma_{\rm min} \lesssim 10$ is needed for a smooth X-ray spectrum, the strongest constraint on $r_{\rm o} / R$ is obtained by using $\gamma_{\rm min} \approx 10$, which implies $n\approx 2$ (taking the minimum to occur at a frequency $10^{17}$\,Hz and a synchrotron peak frequency at $10^{13}$\,Hz). With $\alpha \approx 0.6$ \citep[e.g.,][]{sam09,abd09b}, this shows that $R/r_{\rm o} \gtrsim 20$ gives rise to an X-ray minimum deep enough to be consistent with observations. 

The requirement for a much larger value of $\epsilon_{\rm e}/\epsilon_{\rm B}$ inside the injection region than outside constrains the mechanism whereby the escaping electrons are dispersed throughout the synchrotron emission volume. Adiabatic expansion is unlikely since the value of $\epsilon_{\rm B}$ as well as the radiative efficiency would decrease. A constant value of $B$ and no adiabatic losses of the electrons, as in the above example, are consistent with scattering possibly in connection with streaming. In addition, the acceleration of electrons could occur in regions with low values of $B$, suggesting that the injected electrons may derive their energy from magnetic reconnection \citep[for example, as discussion by][]{nal10}. Both of these processes may take place simultaneously and, hence, the transition from the synchrotron to the inverse Compton component in the spectral energy distribution is sensitive to the combined effects of the acceleration and the subsequent spreading of the electrons through the whole emission volume.

In a MIC-scenario, spectral properties and flux variations are quite different from those in a source where the high frequency emission is mainly due to a single scattering. As discussed above, the actual source properties need not differ too much between the two cases due to the sensitivity of $\gamma_{\rm c}$ to the density of relativistic electrons, in particular, but also to the magnetic field strength/density of external photons. Furthermore, this sensitivity becomes more pronounced for lower values of  $\gamma_{\rm c}$ (cf. eq. [\ref{eq:1.16}]). Even for a continuous distribution of intrinsic source parameters one may then expect a bi-model distribution of observed properties, although transition cases should be observed. The first group could be adequately described by a standard synchrotron/Compton model, in which multiple inverse Compton scatterings may occur but, most importantly, emission in the Klein-Nishina limit would not dominate the radiation energetically. The second group corresponds to the situation discussed above and is characterized by $\gamma_{\rm c}\lesssim \gamma_{\rm abs}$. This latter group would have three distinct observational characteristics: (1) The emitted energy in the IC-component should dominate that in the synchrotron component. (2) With $\gamma_{\rm abs} \sim 10^2$, the peak of $\nu f(\nu)$ in the IC-component should occur at an energy $\lesssim 10^2$ MeV as measured in the rest frame of the source. (3) The optically thin synchrotron flux is emitted by electrons with $\gamma > \gamma_{\rm c}$. Hence, the peak of $\nu f(\nu)$ in the synchrotron component occurs at $\nu \approx \nu_{\rm abs}$.

The properties of these two groups are reminiscent  of those observed for the HSPs (first group) and LSPs (second group). In the simplest version, the main difference between their intrinsic source properties could be a higher value of $\epsilon_{\rm e}/\epsilon_{\rm B}$ in the injection region of LSPs as compared to HSPs. In this scenario, flux variations in the two classes of blazars are expected to have different properties. Since inverse Compton cooling plays a minor role in HSPs, variations in the synchrotron component should correlate with those in the IC-component, while in LSPs such correlations are expected to be more complex (see \S\,5).

\section{Fermi blazars}\label{sect5}

The LAT experiment \citep{atw09} aboard the {\it Fermi Gamma-ray Observatory} has provided a wealth of data related to the high frequency properties of blazars. With a typical bulk Lorentz factor $\Gamma \sim 10$ for the emission region in blazars and emission in the Klein-Nishina limit, the photons detected by LAT are those up-scattered from energies corresponding  roughly to the sensitivity range of the {\it Swift}-BAT instrument. Hence, within the MIC-scenario, conclusions drawn from a comparison of the observed properties of blazars detected both by LAT and BAT should be rather model independent. S09 have discussed the spectral and luminosity correlations for such a sample of blazars. Since the LAT and BAT observations are not simultaneous, a comparison for individual objects is not warranted. However, S09 find that the different subgroups (i.e., LSP, ISP and HSP) have distinctly different spectral properties. 

For the LSPs they find an average LAT spectral index $\alpha_{\rm G} \approx 1.6$ and an average BAT spectral index $\alpha_{\rm X} \approx 0.6$. In the MIC-scenario, LSPs correspond to blazars with $\gamma_{\rm c} \lesssim \gamma_{\rm abs}$. It is therefore likely that the LAT detected emission is due to cooling electrons, which leads to $p=1+\alpha_{\rm G} \approx 2.6$. Since this implies $\alpha_{\rm X} < (p-1)/2$, the X-ray spectrum should be due to electrons with the lowest energies (i.e., $\gamma_{\rm min}$). Two breaks are expected for $\gamma_{\rm min} < \gamma_{\rm c}$, which would result in a rather flat peak in $\nu f(\nu)$. The gamma-ray emission from HSPs is usually interpreted as first order Compton scattering. Although the number of HSPs in the S09 sample is low, their mean value of $\alpha_{\rm G}$ ($\approx 0.8$) is consistent with the same value of $p$ as for the LSPs and negligible cooling. Furthermore, the spectral properties of the ISPs, which fall in between those of LSPs and HSPs, are consistent with being transition cases for which the gamma-ray emission is produced in the Klein-Nishina limit but with the cooling frequency in or somewhat above the LAT energy range. 

\begin{figure}
\epsscale{0.80}
\plotone{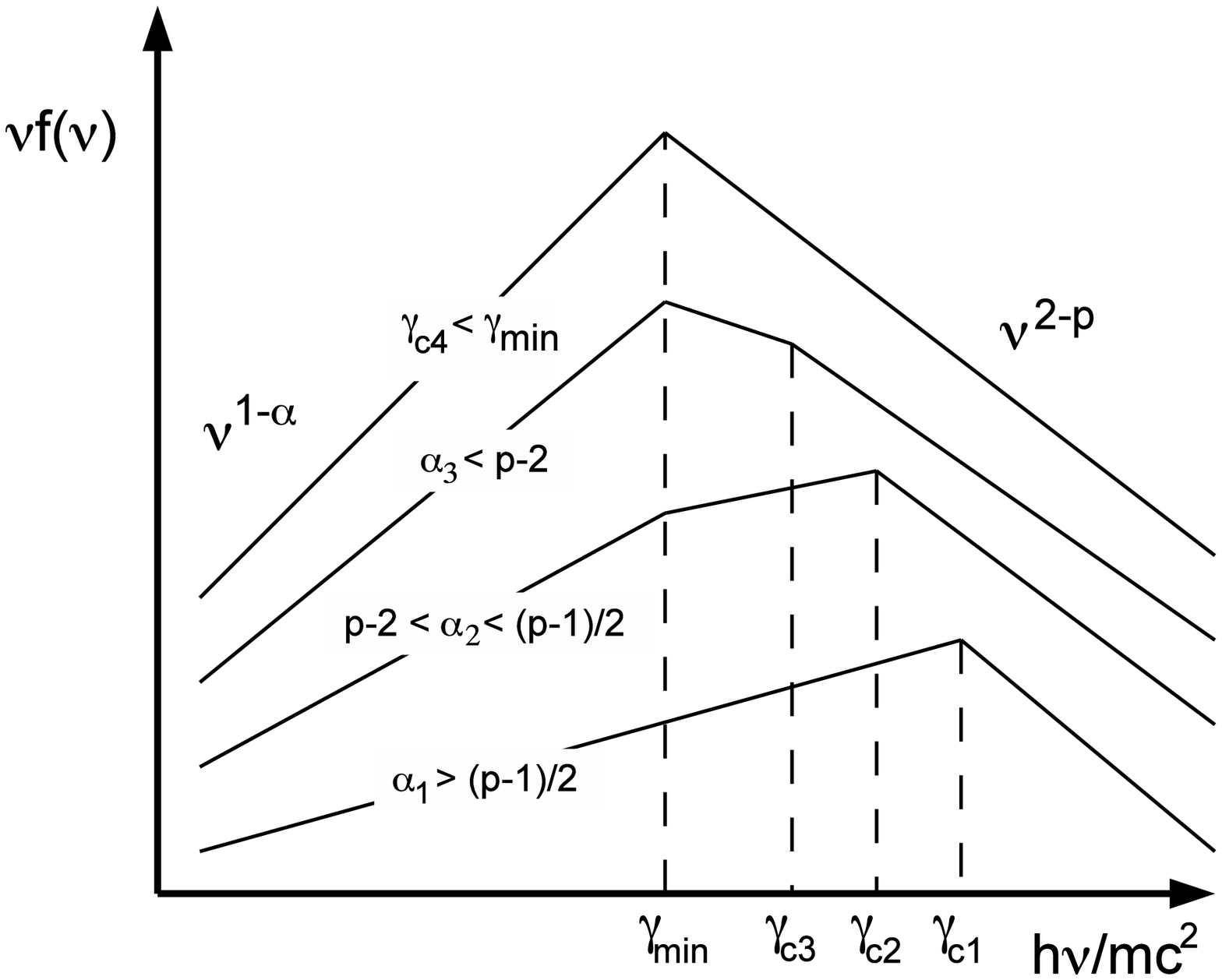}
\caption{Schematic representation of the four qualitatively different energy spectra possible in a multiple scattering scenario when $\alpha < 1$. All the electrons radiate most of their energy in the Klein-Nishina limit. The cases $1-4$ are distinguished by an increasing value of the density of relativistic electrons ($\tau_{\rm o}$). The spectra are labeled both by their value of $\gamma_{\rm c}$ and the corresponding range of values for $\alpha$. {\it Case}\,1: One spectral break; spectral energy peaks at the cooling frequency. {\it Case}\,2: Two spectral breaks; spectral energy peaks at the cooling frequency. {\it Case}\,3: Two spectral breaks; spectral energy peaks at the minimum Klein-Nishina frequency. {\it Case}\,4: One spectral break; spectral energy peaks at the minimum Klein-Nishina frequency.
For $p>3$ only cases $3$ and $4$ are relevant. 
\label{fig1}} 
\end{figure}

It is well known that in general the peak frequency of the IC-component shifts to lower frequencies for higher luminosities \citep[e.g.,][]{fos98}. S09 show that on a more detailed level this is due to more luminous sources having spectra with smaller values of $\alpha_{\rm X}$ as well as larger values of $\alpha_{\rm G}$. Furthermore, this correlation seems to be valid also within the group of the most luminous sources, which have their peak frequencies in between the BAT and LAT frequency ranges. As can be seen from Figure\,\ref{fig1}, this is consistent with a MIC-scenario. 

Hence, the varying spectral properties along the blazar sequence can be interpreted as a continuous variation of the cooling frequency. Although this agrees with the suggestion originally put forward by \citet{fos98}, the implications of this in the MIC-scenario are quite different. The blazar sequence is normally thought to be governed by the energy density of external photons so that a higher density would correspond to a lower cooling frequency and, hence, a higher total luminosity. At some luminosity, the energy density of external photons have decreased to the extent that internally produced synchrotron photons would take over as seed photons \citep[e.g.,][]{bot07}. The former group corresponds to LSPs and the latter to HSPs. LAT observations show that LSPs tend to have $\alpha_{\rm G} > 1$ and ISPs/HSPs $\alpha_{\rm G} \lesssim 1$, with most of the objects with $\alpha_{\rm G} \approx 1$ being classified as ISPs \citep{abd10a}. This indicates that the transition from external photons to internal synchrotron photons as the dominant source of seed photons corresponds to a cooling frequency lying in or close to the LAT frequency range. Such a coincidence remains to be explained in the standard interpretation of the blazar sequence but is expected in the MIC-scenario.

With a bulk Lorentz factor $\Gamma \sim 10$ and emission in the Klein-Nishina limit, the LAT spectral range corresponds, roughly, to electrons with Lorentz factors $\gamma \sim 10^3$. Standard synchrotron theory implies that synchrotron self-absorption sets in for electrons with $\gamma_{\rm abs} \sim 10^2$. In the MIC-scenario, the peak of the synchrotron component in LSPs corresponds, roughly, to the self-absorption frequency. Hence, electrons radiating in the LAT-frequency range have energies similar to those producing the optical synchrotron radiation. The ISPs were originally identified as those blazars having their synchrotron peak frequency in between the LSPs and HSPs (i.e., in the optical/near-UV range). Since this peak frequency corresponds to $\gamma \approx \gamma_{\rm c}$, the cooling frequency in the IC-component for the ISPs is expected to fall in the LAT energy range.

Increasing the energy density of the seed photons (i.e., the external photons and/or the magnetic field) would decrease the cooling frequency. As can be seen from equations (\ref{eq:1.8}) and (\ref{eq:1.9}), this would cause the value of $\alpha$ to either increase ($\alpha > [p-1]/2$) or remain constant ($\alpha < [p-1]/2$). However, S09 show that $\alpha_{\rm X}$ ($\approx \alpha$) actually decreases with increasing luminosity. This implies, in the MIC-scenario, that the main parameter determining the blazar sequence is the density of relativistic electrons (i.e., $\tau_{\rm 0}$). Therefore, the salient features of the blazar sequence can be reproduced by varying only $\tau_{\rm 0}$ (cf. Fig.\,\ref{fig1}). The rough correlation observed between total luminosity and strength of the emission lines in blazars would then not be the primary cause for the blazar sequence but rather a secondary effect induced by a correlation, for example,  between the jet power and the accretion power (i.e., disk and line emission).

The spectral indices used by S09  were calculated assuming a power-law distribution of the flux over the whole LAT energy range \citep{abd09a}. However, as shown by \citet{abd10a}, a spectral steepening at the highest frequencies is common for LSPs and such a behavior is also seen for some ISPs. Although the steepening can be quite strong, it normally occurs at a frequency such that the spectral indices for the lower frequencies are only marginally smaller ($\Delta \alpha_{\rm G} \lesssim 0.1$) than those deduced assuming one power-law for the whole spectral range. These small changes in the value of $\alpha_{\rm G}$ do not significantly change the conclusions of the discussion above; however, the origin of the break for the LSPs remains to be explained. It may be a cooling break akin to the ones in ISPs, in which case a somewhat larger bulk Lorentz factor is implied for LSPs as compared to ISPs. Furthermore, since $p$ is now determined by the spectral index pertaining to the frequency range above the break, its value will be significantly larger than deduced above and, typically, $p \gtrsim 3$ is needed. Such a value is too large to account for the non-cooling spectral  index for the HSPs and, hence, a varying value of $p$ along the blazar sequence is required. 

Another possible cause for the spectral break is absorption on external emission line photons. This process is normally thought to be dominated by Ly\,$\alpha$ photons \citep{g/t09}, in which case the expected break occurs at frequencies too high to be compatible with observations. However, the corresponding resonance line in He\,II can be as strong as Ly\,$\alpha$ in photoionization models with spectra as hard as those in quasars. The absorption then sets in at a frequency a factor of $4$ smaller and would be roughly consistent with the observed breaks in LSPs\footnote{After the submission of the present paper, a similar suggestion was made by Poutanen \& Stern (2010, arXiv:1005.3792)}. The exact break frequency as well as the spectral shape above the break would depend on both the geometry of the emission line region and its location relative the blazar emission region.

The correlations discussed so far have assumed steady state conditions. Such an assumption is not justified during times of rapid changes. Along the blazar sequence, the value of $\alpha_{\rm G}$ increases with blazar luminosity. It has been noted by \citet{ghi09}, that such a correlation is not generally valid for variations in individual blazars; in fact, observations show that in outbursts there is a tendency for variations in the high frequency range to be roughly orthogonal to the one defined by the blazar sequence. Sometimes the high frequency spectral peak even moves from below to above the LAT spectral range, i.e., the spectral properties of LSPs become more ISPs/HSPs like. If the blazar sequence reflects steady state conditions, outbursts could then correspond to variations in the source short enough for cooling effects to become less important in the LAT spectral range.

The much smaller emission region of the inverse Compton component as compared to the synchrotron component would suggest a correspondingly smaller variability times scale as well as flux variations in the former to precede those in the latter. However, this assumes that the variability timescale is long enough for steady state conditions to be reached in the inverse Compton emission region. As just discussed, there are indications that rapid outbursts in LSPs in a MIC-scenario occur under non-steady state conditions resulting in less severe inverse Compton cooling. The fraction of the injected energy emitted as synchrotron emission would therefore increase. Furthermore, a substantial part of the synchrotron emission could be emitted within the injection region leading to similar variability times scales for the two flux components as well as a considerably smaller time lag. Hence, if the correlated time variations sometimes seen in the synchrotron and inverse Compton components are due to non-steady state conditions, they are expected to be accompanied by changes in the high frequency range, which is not along the blazar sequence. Another effect that could smear out the expected differences between the variability properties of the two flux components, even under steady state conditions, is the structure of the injection region. It has been tacitly assumed that the injection of energy occurs in a single region. However, it could take place in several regions located throughout the synchrotron emission volume; for example, a quasi-homogeneous distribution of injection regions should result in similar variability properties for the synchrotron and inverse Compton components.

\section{Conclusions}\label{sect6}

Multiple scatterings give rise to observed properties quite different from those expected in a single scattering scenario. This is mainly due to the different relations between observed photon frequency and electron energy; for example, in the case of multiple scatterings, the emission over a wide frequency range can be produced by electrons with the same energy, while in the single scattering case, there is a direct correlation between emitted frequency and electron energy. With respect to the general properties of multiple scattering scenarios, the main points of the present paper are:

\noindent $\bullet$ The description of the multiple scattering scenario for a thermal/mono-energetic distribution of electrons is extended to also include a power-law distribution.

\noindent $\bullet$ A power-law distribution of electron energies introduces an extra spectral break. 

\noindent $\bullet$ The various spectral indices are determined not only by the slope of the power-law distribution but also by the density of electrons; hence, spectral differences do not necessarily indicate intrinsically different electron distributions.

\noindent $\bullet$ The correlation of time variations in the IC-component with those in a low-frequency synchrotron component is expected to change character when cooling becomes important for electrons radiating close to or below the synchrotron self-absorption frequency. 

When the observed emission from blazars is interpreted within a multiple scattering scenario, the deduced source properties differ in some respects from those derived from a single scattering scenario. The main implications from a multiple scattering scenario are:

\noindent $\bullet$ The rate of radiative cooling is more sensitive to the density of seed photons than for single scatterings. Even with a rather homogeneous distribution of intrinsic source properties, this non-linearity suggests that most blazars would belong to one of two groups, characterized either by (i) cooling in the multiple scattering regime is not important or (ii) multiple scatterings cause cooling to be important even for electrons radiating close to or below the synchrotron self-absorption frequency. This bimodal distribution is reminiscent of the blazar classes HSP and LSP, respectively.

\noindent $\bullet$ When multiple scatterings are energetically important, the high energy radiation is emitted in the Klein-Nishina limit. Electrons radiating in the {\it Fermi}-LAT frequency range would then also emit synchrotron radiation in the optical/near-UV range. This implies that blazars with their synchrotron cooling frequency in the optical/near-UV should have a corresponding cooling break in the {\it Fermi}-LAT frequency range. The ISP blazars have properties consistent with being such transition objects.

\noindent $\bullet$ The observed anti-correlation between the total luminosity of the IC-component and the spectral index in the X-ray/{\it Swift}-BAT range is best understood as resulting from differences in the energy density of electrons rather than that of the seed photons. In fact, it is possible that the whole blazar sequence is primarily driven by variations in the energy density of electrons.

\noindent $\bullet$ One-zone models are not adequate to describe the spectral transition region between the synchrotron and IC components in blazars. Instead a source structure is suggested where injection of relativistic electrons occurs in a small fraction of the synchrotron emission volume. The IC-component is produced mainly within the injection region(s), while the synchrotron emission is due to electrons, which have escaped from the injection region(s). Such a source structure preserves the main spectral correlations expected between the synchrotron and IC-components in one-zone models.

\acknowledgements

This research was supported by a grant from the Swedish Natural Science Research Council. It was done as part of the International Team collaboration number 160 sponsored by the International Space Science Institute (ISSI) in Switzerland.

\clearpage

\end{document}